\documentclass[9pt,twocolumn,twoside]{opticajnl}
\journal{opticajournal} % use for journal or Optica Open submissions

\setboolean{shortarticle}{true}
\usepackage{pdfpages} % for \includepdf
\usepackage{lineno}
\DeclareMathOperator{\re}{Re}

\newcommand{\oma}{\omega_\mathrm{A}}

\newcommand{\oml}{\omega_\mathrm{L}}
\newcommand{\taud}{\tau_\mathrm{d}}

\title{Signatures of inter-sideband coherence in the resonance fluorescence spectrum of an acoustically-modulated quantum dot}

\author[1]{Rafał A. Bogaczewicz}
\author[2]{Hubert J. Krenner}
\author[1,*]{Paweł Machnikowski} % Tu wymuszamy, aby gwiazdka i jedynka były razem

\affil[1]{Institute of Theoretical Physics, Wrocław University of Science and Technology, 50-370 Wrocław, Poland}
\affil[2]{Institute of Physics, University of Münster, 48149 Münster, Germany}
\affil[*]{pawel.machnikowski@pwr.edu.pl} % Tradycyjny mail jako osobna "afiliacja"

\begin{abstract}
We theoretically investigate the inter-sideband phase coherence within the resonance fluorescence spectrum of an acoustically modulated quantum dot using a filtered-field formalism for a Mach--Zehnder configuration. We demonstrate that geometric slant of the interferograms provides an indicator of phase coherence that is resilient to environmental white noise. Specifically, noise-induced spectral diffusion reduces the global fringe intensity, while leaving the characteristic inclination strictly invariant. Our findings establish a framework for verifying single-photon coherence between spectral sidebands, essential for frequency-bin encoding and scalable quantum networking in realistic, noisy solid-state architectures.
\end{abstract}

\setboolean{displaycopyright}{false} % Do not include copyright or licensing information in submission.

\begin{document}

\maketitle

\section{Introduction} \label{sec:Introduction}

Quantum hybrid systems \cite{Xiang2013,Kurizki2015} enable advanced photonic state engineering by integrating diverse physical platforms \cite{Clerk2020}. For instance, in acousto-optic on-chip devices, surface acoustic waves (SAWs) \cite{Schuetz2015,Krenner2026} modulate the transition energy of a quantum dot (QD) at GHz frequencies, making the QD an efficient acousto-mechanical interface for resonance fluorescence (RF). In the resolved sideband regime, the RF spectrum consists of a central line at laser frequency \cite{ScullyZubairy1997} and a series of acoustically induced sidebands \cite{Metcalfe2010,Villa2017}. The single-photon character of this radiation, confirmed by photon antibunching \cite{Villa2017,Weiss2021}, ensures that sideband intensities map directly to single-photon scattering probabilities into specific frequency channels. Tuning the acoustic mode allows for control in both frequency \cite{Weiss2021} and time \cite{Wigger2021} domains, forming the basis for time- and frequency-bin encoding \cite{Pan2012,Lu2023}, quantum transduction \cite{Stannigel2010}, and multiplexing \cite{Piparo2019}. However, the fundamental question regarding the coherence between these spectral sidebands remains open.

In state-of-the-art implementations \cite{Wan2022,Descamps2024}, merely shifting the photon energy into a sideband is insufficient for full quantum control because the efficiency of energy transfer must be accompanied by the preservation of quantum coherence. Maintaining a well-defined phase relationship between spectral components is essential to deterministically create frequency-bin superpositions. Without a fixed phase relation, the generated states lose coherence and revert to a classical regime of incoherent scattering. 

Moreover, in realistic semiconductor environments, coherence is constantly challenged by stochastic charge \cite{Kuhlmann2013} or spin \cite{Kasprzak2022} fluctuations, which lead to emitter transition energy diffusion \cite{Bogaczewicz2023} and fundamentally alter the emission spectrum. Specifically, such phase diffusion gives rise to two coexisting series of spectral lines \cite{Bogaczewicz2025}: an elastic component that remains unbroadened, located at the laser frequency (and its acoustic multiples), and an inelastic component centered at the emitter's transition frequency, which undergoes systematic homogeneous broadening according to the noise strength. While recent studies have extensively analyzed how such environmental dephasing limits the performance of resonantly driven quantum emitters, focusing primarily on the resulting loss of visibility \cite{Matthiesen2012}, its precise impact on the mutual coherence between distinct spectral sidebands remains unexplored.

In this Letter, we theoretically investigate the mutual coherence between sidebands in the RF spectrum of an acoustically modulated system using the filtered-field formalism \cite{Eberly1977}. We model the interference between these components in a Mach–Zehnder configuration with a relative frequency offset and a time-delay-induced optical phase. We show that the interband coherence is directly encoded in a qualitative feature, a slant, of the resulting time-dependent interferogram in the time-phase plane. We prove that this tilt serves as a robust indicator of inter-sideband coherence: While the white noise changes the overall intensity of the signal, it leaves the geometric slant invariant, establishing the proposed interferometry as a reliable test of single-photon coherence between frequency channels.

\section{Model and formalism} \label{sec:Model_formalism}

\begin{figure}[tb]
\includegraphics[width=\linewidth]{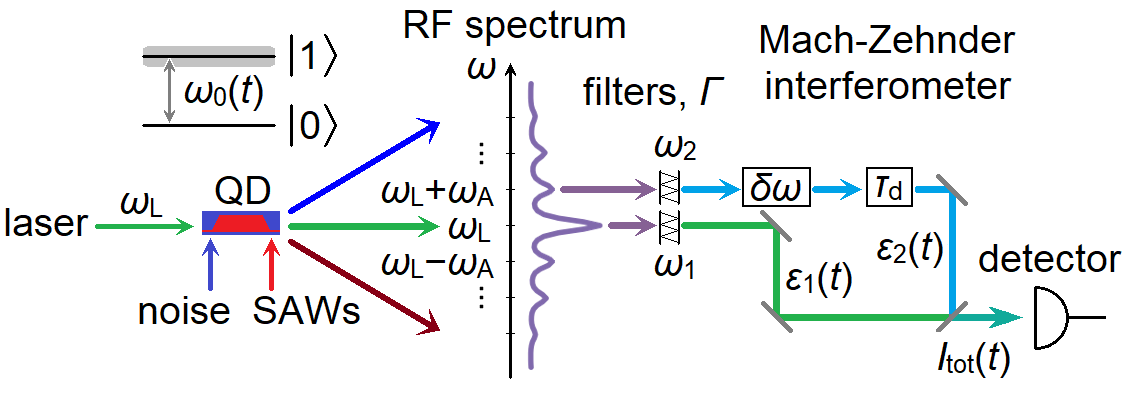}
\caption{\label{fig:system} Schematic representation of the theoretical setup. Left: A QD with a time-modulated transition frequency (indicated by the shaded area). Center: Two Lorentzian filters with bandwidth $\Gamma$ are tuned to selected sidebands in the output signal. Right: The selected beams enter a Mach-Zehnder interferometer. One arm incorporates an external frequency shift $\delta\omega$ and a time delay $\taud$.}
\end{figure}

We consider a self-assembled semiconductor QD driven by a weak, monochromatic laser field with frequency $\oml/(2\pi)\sim 10^2$ THz (see Fig.~\ref{fig:system}). The emitter is modeled as a two-level system (TLS) with ground ($|0\rangle$) and excited ($|1\rangle$) states. This simplification is well-justified by optical selection rules under circularly polarized excitation \cite{Machnikowski2008}. The system undergoes spontaneous emission at a rate $\gamma$ (in typical InGaAs QDs $\gamma\approx 1$ GHz \cite{Regelman2001}).

The transition frequency of the QD is
\begin{equation}
\omega_0(t) = \omega_0 + \Delta\omega_{\text{ac}}(t) + \Delta\omega_{\text{ns}}(t),
\end{equation}
where $\omega_0/(2\pi)\sim 10^2$~THz is the unperturbed frequency, $\Delta\omega_{\text{ac}}(t)$ denotes the periodic modulation by surface acoustic waves (SAWs) \cite{Weiss2021, Wigger2021} with fundamental frequency $\oma/(2\pi)\sim 1$~GHz and $\Delta\omega_{\text{ns}}(t)$ represents the contribution of rapid environmental fluctuations \cite{Bogaczewicz2025}. The stochastic term is modeled as a white noise process $\langle \Delta\omega_{\text{ns}}(t)\Delta\omega_{\text{ns}}(t+\tau) \rangle = 2D\delta(\tau)$, where $D$ is the phase diffusion coefficient. The first-order correlation function of the scattered light $G(t_\mathrm{a}, t_\mathrm{b}) = \langle \sigma_+(t_\mathrm{a})\sigma_-(t_\mathrm{b}) \rangle$, incorporating the acousto-optic driving mechanism, noise, and the Lindblad dissipation, is calculated along the lines developed in our previous works \cite{Weiss2021,Wigger2021,Bogaczewicz2025}.

To analyze the mutual coherence between different spectral lines, we apply the time-dependent filtered-field formalism \cite{Eberly1977} to a Mach–Zehnder interferometer. The field operators at the two input channels of the final beam splitter are defined as:
\begin{subequations} \label{eq:eps}
\begin{align}
\varepsilon_1(t) &= \int_{-\infty}^{\infty} ds F_1(s)\sigma_-(t-s), \label{eq:eps_1} \\
\varepsilon_2(t) &= e^{i\delta\omega t} \int_{-\infty}^{\infty} ds F_2(s)\sigma_-(t - \taud - s), \label{eq:eps_2}
\end{align}
\end{subequations}
where $F_j(s) = \Gamma \exp(-i\omega_j s - \Gamma s)\theta(s)$ is the temporal response of a Lorentzian filter with bandwidth $\Gamma$ centered at frequency $\omega_j$. The first arm acts as a reference arm, whereas the second arm incorporates an external frequency shift $\delta\omega$ and a time delay $\taud$.

The total intensity recorded at the detector at time $t$,
\begin{equation}
I_{\text{tot}}(t) = \langle [\varepsilon_1^\dagger(t) + \varepsilon_2^\dagger(t)][\varepsilon_1(t) + \varepsilon_2(t)] \rangle , \label{eq:I_tot}
\end{equation}
splits into individual filter intensities and an interference term, $I_{\text{tot}}(t) = I_1(t) + I_2(t) + I_{\text{int}}(t)$. By substituting Eqs.~(\ref{eq:eps}) and using the definition of the autocorrelation function  (see the Supplement), we get
\begin{subequations}
\begin{align}
I_{1,2}(t) = & \Gamma^2 \int_0^\infty ds \int_0^\infty ds' e^{-\Gamma(s+s')+i\omega_{1,2}(s-s')} \label{eq:I_1} \\
& \quad\times  G(t_{1,2}-s,t_{1,2}-s'), \nonumber \\
I_\mathrm{int}(t) = & 2\Gamma^2 \re  e^{i\delta \omega t} \int_0^\infty ds \int_0^\infty ds' e^{-\Gamma(s+s')+i(\omega_1 s - \omega_2s')} \label{eq:I_int} \\
& \quad \times G(t_1-s,t_2-s'), \nonumber
\end{align}
\end{subequations}
where $t_1=t$ and $t_2 = t-\taud$.

We assume that $t$ is long enough to reach the stationary state and $\omega_0,\oml,\omega_1,\omega_2\gg \oma,|\oml-\omega_0|,\gamma,D,\Gamma$. 

There are two signatures of coherence that can be exploited in different experimental setups. One approach is to frequency-shift one of the sidebands to align frequencies and observe a standard Mach-Zehnder interference in a time-integrated signal. The other is to superpose two sidebands directly and observe beats at the frequency difference. The latter is less demanding experimentally, excluding the need for an additional acousto-optic modulator or similar equipment to shift the frequency. However, since the intensity of the sideband oscillates itself \cite{Wigger2021}, coherence-related beats may be difficult to separate. We therefore propose a unified approach based on two-dimensional interferograms with real time and phase shift on the two axes, in which the frequency shift is a parameter.

\section{Results} \label{sec:Results}

We present two-dimensional interferograms, which present the total intensity $I_{\mathrm{tot}}(t)$ as a function of real time $t$ (in units of $1/\oma$) and time-delay-induced optical phase $\oml \taud$, for the acoustically modulated QD using the parameter ratios $\gamma/\oma = 2\Gamma/\oma = 0.1$ and a single SAW mode $\Delta\omega_{\mathrm{ac}}(t) = 2\oma \cos(\oma t)$. The spectral filters (see Fig.~\ref{fig:system}) select lines of the RF spectrum with  $\omega_{1,2} = \oml + k_{1,2}\oma$ for an integer $k_{1,2}$. We normalize the signals to the intensity of standard, unperturbed RF, $I_0 = \pi \Omega^2 / \gamma^2$ \cite{ScullyZubairy1997}.

\begin{figure}[t]
\centering
\includegraphics[width=\linewidth]{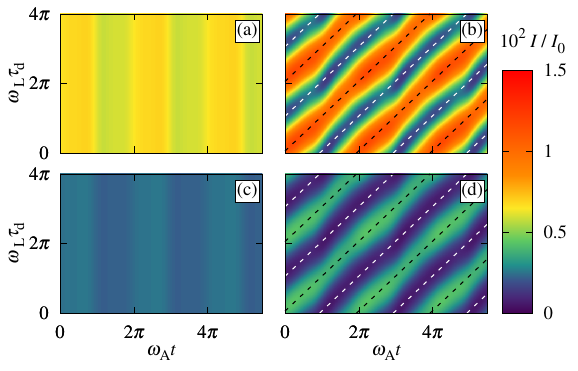}
\caption{Incoherent (a, c) and coherent (b, d) superpositions of the light beams from the ZPL ($k_1=0$) and the first PSB ($k_2=1$). Panels (a, b) show the noise-free case ($D=0$), while (c, d) incorporate white noise ($D = 0.05\oma$). Here $\oml = \omega_0$ and $\delta\omega = 0$. Black and white dashed lines in (b, d) represent the analytical trajectories for maximal and minimal intensities, respectively.}
\label{fig:interferograms_coherent_incoherent}
\end{figure}

We first investigate the interference between the central (``zero-phonon'') line (ZPL, $k_1 = 0$) and the first phonon sideband (PSB, $k_2 = 1$). Fig.~\ref{fig:interferograms_coherent_incoherent} presents the resulting interferograms. In the purely incoherent regime (Fig.~\ref{fig:interferograms_coherent_incoherent}a), the total intensity simplifies to the optical-phase-independent sum $I_1(t)+I_2(t)$. Since the lack of coherence excludes any phase-dependence, the pattern consists of purely vertical modulations, reflecting the individual sideband oscillations under acoustical modulation \cite{Wigger2021}. In contrast, when phase coherence is preserved (Fig.~\ref{fig:interferograms_coherent_incoherent}b), phase-dependence appears, which turns out to take the form of a systematic slant of the fringes. The non-uniform thickness and envelope modulation of these fringes stem from the intensity modulation, which can be traced back to beatings between neighboring spectral components covered by the finite filter bandwidth \cite{Wigger2021} (see also Supplement). When environmental white noise is introduced (Figs.~\ref{fig:interferograms_coherent_incoherent}c,d), the qualitative structure remains the same: although fluctuations reduce the overall intensity, the geometric slant is preserved, confirming the phase coherence of scattered photons.

The slant of the interference fringes is determined by the phase-matching condition in the ideal filter limit $\Gamma \to 0$ (see the Supplement for the full derivation),
\begin{equation}
\oml \taud = N\pi - \left( k_1 - k_2 + \frac{\delta\omega}{\oma} \right) \oma t - \theta_{k_1, k_2}, \label{eq:geometric_slant}
\end{equation}
where even and odd integers $N$ correspond to intensity maxima and minima, respectively, and the phase shift $\theta_{k_1, k_2}$ depends on the sideband orders, $k_1$ and $k_2$ (see the Supplement). As shown in Figs.~\ref{fig:interferograms_coherent_incoherent}b and d, the dashed lines representing Eq.~(\ref{eq:geometric_slant}) perfectly track the fringes in both regimes  (here $\theta_{0,1} \approx -0.181$). \eqref{eq:geometric_slant} directly maps the sideband separation onto the fringe inclination. 

\begin{figure}[t]
\centering
\includegraphics[width=\linewidth]{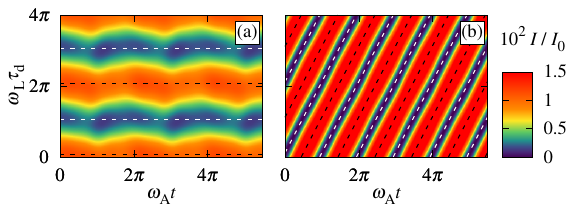}
\caption{Coherent interference intensities without noise ($D=0$) and under resonant ($\oml = \omega_0$) driving. (a) Superposition of the ZPL and the first right PSB, $(k_1, k_2) = (0, 1)$, under the resonance shift condition $\delta\omega = \oma$. (b) Superposition of the left and right first-order PSBs, $(k_1, k_2) = (-1, 1)$, without an external shift ($\delta\omega = 0$). Black and white dashed lines represent the analytical trajectories for maximal and minimal intensities, respectively.}
\label{fig:slant_control}
\end{figure}

To further illustrate the behavior of the interferogram geometry, we examine the system response for two other cases under different sideband configurations and external frequency shifts $\delta\omega$. Fig.~\ref{fig:slant_control} presents a comparison highlighting how internal mode selection and external modulation cooperate to dictate the fringe behavior.

We first explore the case where the external frequency shift $\delta\omega=\oma$ eliminates the time-dependent beating between fields of different frequencies for the ZPL–PSB configuration, $(k_1, k_2) = (0, 1)$. The resulting interferogram is shown in Fig.~\ref{fig:slant_control}a. Frequency matching eliminates the beats and removes the interferogram time dependence, causing the slope term in the parentheses of Eq.~(\ref{eq:geometric_slant}) to vanish completely, thereby reducing the system to a standard Mach-Zehnder interference scenario. Consequently, the interference fringes align horizontally, becoming nearly independent of time. As shown by the dashed lines, Eq.~(\ref{eq:geometric_slant}) perfectly tracks these horizontal structures of intensity minima and maxima, respectively. In this regime, the recorded intensity is quasi-stationary within the acoustic cycle (apart from minor envelope beating from the finite filter bandwidth $\Gamma$) and its global structure is governed entirely by the optical phase delay $\oml \taud$. The interference pattern would also appear in this case in the time integrated signal. 

For comparison, Fig.~\ref{fig:slant_control}b shows the interference between the left and right first-order sidebands, $(k_1, k_2) = (-1, 1)$, in the absence of an external shift ($\delta\omega = 0$). In this configuration, the phase-matching slope in Eq.~(\ref{eq:geometric_slant}) is determined solely by the discrete mode spacing ($k_1 - k_2 = -2$), resulting in strongly inclined fringes. The corresponding analytical lines (dashed), evaluated using the phase $\theta_{-1,1} \approx 2.779$, again perfectly match the full numerical simulation. This shows that tuning $\delta\omega$ and selecting various spectral lines allows for direct control over the interferometric signatures of phase relationship between the selected spectral components.

\begin{figure}[t]
\centering
\includegraphics[width=\linewidth]{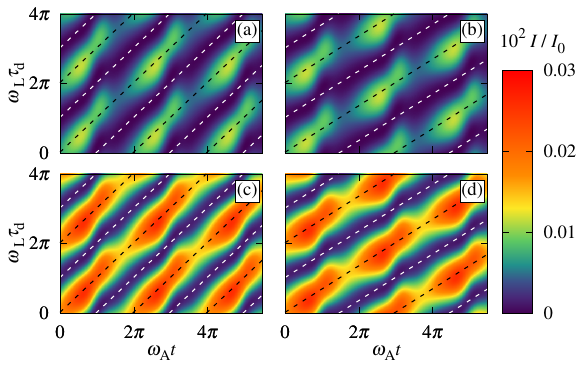}
\caption{Coherent interference intensities for $\oml = \omega_0 + \oma/3$ under the $(k_1,k_2)=(0,1)$ configuration. Panels (a, c) show $\delta\omega=0$, while (b, d) show $\delta\omega=\oma/3$. Results for the noise-free case (a, b) and $D = 0.05\oma$ (c, d). Black and white dashed lines denote the analytical trajectories for maximal and minimal intensities, respectively.}
\label{fig:detuned_case}
\end{figure}

To evaluate the operational robustness of the proposed interferometric framework against laser detuning and environmental noise, in Fig.~\ref{fig:detuned_case} we analyze the off-resonant excitation regime ($\Delta = \oml - \omega_0 = \oma/3$) for $(k_1, k_2) = (0, 1)$ under varying frequency shifts $\delta\omega$ and in the absence or presence of noise.  Due to the off-resonant driving, the scattered light intensity in all cases is significantly suppressed compared to the resonant regime (note the rescaled colorbar). Under detuned excitation, the detailed form of the interferogram changes visibly, confining the high-intensity areas to a pattern of spots (note, however, that the exact pattern depends on the temporal width of the spectral filter $1/\Gamma$). Crucially, even in this case, the geometric slant is perfectly readable from the interferogram and aligns with the analytical axes calculated from Eq.~(\ref{eq:geometric_slant}) (dashed lines, here $\theta_{0,1}\approx -0.076$).
With environmental noise ($D > 0$, Figs.~\ref{fig:detuned_case}c,d), the overall intensity grows, which is a consequence of approaching the laser frequency by the fluctuating transition energy. Moreover, the slanted fringes become slightly more uniform (the contrast along the black dashed lines drops by about 1/3). Again, the geometric slant remains clear and follows Eq.~(\ref{eq:geometric_slant}). This proves that the acousto-optic phase information is structurally protected against fast noise, demonstrating that the interferometric fringe inclination provides a highly resilient signature of inter-subband coherence, even when the emitter undergoes noise-induced dephasing. On the other hand, inhomogeneous broadening, i.e., integrating the signal over multiple repetitions of the experiment with the transition energy shifted due to slowly changing environmental conditions, affects the interferogram via detuning-dependent phase shifts $\theta_{k_1,k_2}$ (see the Supplement), which has limited effect as long as the standard deviation of the energy fluctuation (the inhomogeneous line width) is well below the acoustic frequency. Moreover, the contribution of strongly detuned spectra, $\Delta\gg\gamma$, is small due to the Lorentzian excitation profile, so that the inhomogeneity effect is suppressed if $\gamma \ll \oma$. From this perspective, systems allowing much higher acoustic frequencies, including those based on bulk acoustic waves \cite{Groll2026,Kuznetsov2026}, may be favorable here. 

\section{Conclusions} \label{sec:Conclusions}

We have established a robust protocol for characterizing the interband coherence of fluorescence photons in an acoustically modulated QD. Our framework leverages the slanted layout of two-dimensional time-dependent interferograms to provide a direct verification of a fixed phase relation between light emitted to different modulation-induced sidebands. The proposed protocol uses a frequency offset of one of the beams as a parameter and thus combines the observation of Mach-Zehnder interference with time-domain beating between different frequency sidebands, comprising interferometric evidence that may be observable in different kinds of experimental setups. 
In particular, adjusting the parametric frequency offset to the frequency difference between the sidebands leads to standard Mach-Zehnder interferometry, where an interferogram also emerges in the time-integrated signal.
In this way, our theoretical analysis provides a tool for the assessment of the coherent functionality of this elementary acousto-optic device. In further perspective, such an assessment of coherence may be critical for frequency-bin encoding protocols, where a rigid and stable optical phase relationship across independent spectral channels is mandatory to form high-fidelity qubit superpositions.

The proposed approach is resilient to environmental fluctuations, which might be expected to degrade the visibility of interferometric fringes in a realistic solid-state environment. Here, however, the slant of the cross-sideband interference fringes remains visible and does not change in the presence of Gaussian white noise, representing fast environmental fluctuations that lead to dephasing. In addition, the pattern is, to some extent, resilient to inhomogeneous broadening effects resulting from slow noise. As an additional test of control over the optical coherence, varying the parametric frequency shift allows one to precisely manipulate the phase dependence, inducing a continuous rotation of the interference fringes. 

The proposed time-domain interferometric technique relies on the enormous phase stability of acoustic modulation \cite{Weiss2021} and does not require active feedback or stabilization to extract phase relationships, providing a practical diagnostic tool to confirm the quantum coherence of solid-state components in realistic, noisy semiconductor environments.

\begin{backmatter}

\bmsection{Funding} P.M. and R.A.B. acknowledge support by the Polish National Science Centre (Narodowe Centrum Nauki, NCN) under Grant No. 2023/50/A/ST3/00511). H.J.K. acknowledges support by the Deutsche Forschungsgemeinschaft (DFG, German Research Foundation), projects 465136867, 505596454 The authors are grateful to Alexander von Humboldt-Stiftung for support via a research group linkage grant.

%\bmsection{Acknowledgment} 

\bmsection{Disclosures} The authors declare no conflicts of interest.

\bmsection{Data Availability Statement} No data were generated or analyzed in the presented research.

\bmsection{Supplemental document} See Supplement 1 for supporting content.

\end{backmatter}

\bibliography{publikacja}

\begin{thebibliography}{10}
\newcommand{\enquote}[1]{``#1''}

\bibitem{Xiang2013}
Z.-L. Xiang, S.~Ashhab, J.~Q. You, and F.~Nori, {\protect\JournalTitle{Rev. Mod. Phys.}} \textbf{85}, 623 (2013).

\bibitem{Kurizki2015}
G.~Kurizki, P.~Bertet, Y.~Kubo, \emph{et~al.}, {\protect\JournalTitle{Proceedings of the National Academy of Sciences}} \textbf{112}, 3866 (2015).

\bibitem{Clerk2020}
A.~A. Clerk, K.~W. Lehnert, P.~Bertet, \emph{et~al.}, {\protect\JournalTitle{Nature Physics}} \textbf{16}, 257 (2020).

\bibitem{Schuetz2015}
M.~J.~A. Schuetz, E.~M. Kessler, G.~Giedke, \emph{et~al.}, {\protect\JournalTitle{Phys. Rev. X}} \textbf{5}, 031031 (2015).

\bibitem{Krenner2026}
H.~J. Krenner, P.~V. Santos, C.~Westerhausen, \emph{et~al.}, {\protect\JournalTitle{Journal of Physics D: Applied Physics}} \textbf{59}, 093001 (2026).

\bibitem{ScullyZubairy1997}
M.~O. Scully and M.~S. Zubairy, \emph{Quantum Optics} (Cambridge University Press, 1997).

\bibitem{Metcalfe2010}
M.~Metcalfe, S.~M. Carr, A.~Muller, \emph{et~al.}, {\protect\JournalTitle{Phys. Rev. Lett.}} \textbf{105}, 037401 (2010).

\bibitem{Villa2017}
B.~Villa, A.~J. Bennett, D.~J.~P. Ellis, \emph{et~al.}, {\protect\JournalTitle{Applied Physics Letters}} \textbf{111}, 011103 (2017).

\bibitem{Weiss2021}
M.~Weiß, D.~Wigger, M.~Nägele, \emph{et~al.}, {\protect\JournalTitle{Optica}} \textbf{8}, 291 (2021).

\bibitem{Wigger2021}
D.~Wigger, M.~Weiß, M.~Lienhart, \emph{et~al.}, {\protect\JournalTitle{Physical Review Research}} \textbf{3}, 033197 (2021).

\bibitem{Pan2012}
J.-W. Pan, Z.-B. Chen, C.-Y. Lu, \emph{et~al.}, {\protect\JournalTitle{Reviews of Modern Physics}} \textbf{84}, 777 (2012).

\bibitem{Lu2023}
H.-H. Lu, M.~Liscidini, A.~L. Gaeta, \emph{et~al.}, {\protect\JournalTitle{Optica}} \textbf{10}, 1655 (2023).

\bibitem{Stannigel2010}
K.~Stannigel, P.~Rabl, A.~S. S\o{}rensen, \emph{et~al.}, {\protect\JournalTitle{Phys. Rev. Lett.}} \textbf{105}, 220501 (2010).

\bibitem{Piparo2019}
N.~Lo~Piparo, W.~J. Munro, and K.~Nemoto, {\protect\JournalTitle{Phys. Rev. A}} \textbf{99}, 022337 (2019).

\bibitem{Wan2022}
L.~Wan, Z.~Yang, W.~Zhou, \emph{et~al.}, {\protect\JournalTitle{Light: Science {\&} Applications}} \textbf{11}, 145 (2022).

\bibitem{Descamps2024}
T.~Descamps, T.~Schetelat, J.~Gao, \emph{et~al.}, {\protect\JournalTitle{Nano Letters}} \textbf{24}, 12493 (2024).

\bibitem{Kuhlmann2013}
A.~V. Kuhlmann, J.~Houel, A.~Ludwig, \emph{et~al.}, {\protect\JournalTitle{Nature Physics}} \textbf{9}, 570 (2013).

\bibitem{Kasprzak2022}
J.~Kasprzak, D.~Wigger, T.~Hahn, \emph{et~al.}, {\protect\JournalTitle{ACS Photonics}} \textbf{9}, 1033 (2022).

\bibitem{Bogaczewicz2023}
R.~A. Bogaczewicz and P.~Machnikowski, {\protect\JournalTitle{New Journal of Physics}} \textbf{25}, 093057 (2023).

\bibitem{Bogaczewicz2025}
R.~A. Bogaczewicz and P.~Machnikowski, {\protect\JournalTitle{Opt. Lett.}} \textbf{50}, 888 (2025).

\bibitem{Matthiesen2012}
C.~Matthiesen, A.~N. Vamivakas, and M.~Atat\"ure, {\protect\JournalTitle{Phys. Rev. Lett.}} \textbf{108}, 093602 (2012).

\bibitem{Eberly1977}
J.~H. Eberly \emph{et~al.}, {\protect\JournalTitle{Journal of the Optical Society of America}} \textbf{67}, 1252 (1977).

\bibitem{Machnikowski2008}
P.~Machnikowski, A.~Grodecka, C.~Weber, and A.~Knorr, {\protect\JournalTitle{Materials Science-Poland}} \textbf{26}, 829 (2008).

\bibitem{Regelman2001}
D.~V. Regelman, U.~Mizrahi, D.~Gershoni, \emph{et~al.}, {\protect\JournalTitle{Phys. Rev. Lett.}} \textbf{87}, 257401 (2001).

\bibitem{Groll2026}
D.~Groll, D.~Wigger, M.~Wei{\ss}, \emph{et~al.}, {\protect\JournalTitle{J. Phys. Photonics}} \textbf{8}, 012008 (2026).

\bibitem{Kuznetsov2026}
A.~S. Kuznetsov, M.~Saeedi, Z.~Wang, \emph{et~al.}, preprint arXiv:2510.22826 (2026).

\end{thebibliography}

\clearpage


\section*{Supplementary material}

# Signatures of interband coherence in the resonance fluorescence spectrum of an acoustically-modulated quantum dot

This is the supplementary material for the article [Link].

## CONTENTS



## 1. INTENSITIES

In this section we derive the formulas for the intensities from each beam and the interference term between them.

We start from Eq. (3) from the main text, which gives us the three components of the intensity

$$I_1(t) = \langle \varepsilon_1^\dagger(t)\varepsilon_1(t)\rangle, \qquad I_2(t) = \langle \varepsilon_2^\dagger(t)\varepsilon_2(t)\rangle, \qquad I_{\text{int}}(t) = \langle \varepsilon_1^\dagger(t)\varepsilon_2(t) + \text{h.c.}\rangle. \tag{S1}$$

Then we use Eqs. (2) and the definition of the autocorrelation function from the main text, obtaining

$$I_1(t) = \int_{-\infty}^{\infty} ds \int_{-\infty}^{\infty} ds' F_1^*(s)F_1(s')G(t-s, t-s'), \tag{S2a}$$

$$I_2(t) = \int_{-\infty}^{\infty} ds \int_{-\infty}^{\infty} ds' F_2^*(s)F_2(s')G(t-\tau_{\text{d}}-s, t-\tau_{\text{d}}-s'), \tag{S2b}$$

$$I_{\text{int}}(t) = 2\,\text{Re}\, e^{i\delta\omega t} \int_{-\infty}^{\infty} ds \int_{-\infty}^{\infty} ds' F_1^*(s)F_2(s')G(t-s, t-\tau_{\text{d}}-s'). \tag{S2c}$$

As shown in Ref. [1], in the presence of white noise, the autocorrelation function contains the elastic (el) and inelastic (inel) parts, which are

$$G_{\text{el}}(t_1,t_2) = \frac{\Omega^2}{\gamma^2} e^{-i\omega_{\text{L}}(t_2-t_1)} \sum_{n,m} c_n c_m^* e^{i\omega_{\text{A}}(nt_1 - mt_2)} \quad \text{for every } t_1, t_2, \tag{S3a}$$

$$G_{\text{inel}}(t_1,t_2) = \frac{\Omega^2}{2\gamma^2} e^{-\left(\frac{\gamma}{2}+D-i\Delta+i\omega_{\text{L}}\right)(t_2-t_1)} \sum_{n,m,k} b_m^* b_{n-k} d_k e^{i\omega_{\text{A}}(nt_1 - mt_2)} \quad \text{for } t_1 < t_2. \tag{S3b}$$

Rabi frequency and detuning are defined, respectively, as $\Omega$ and $\Delta = \omega_{\text{L}} - \omega_0$. The coefficients $b_n, c_n$ and $d_n$ for the acoustic modulation assumed in the main text are [1]

$$b_n = J_n(A),$$

$$c_n = \frac{\gamma i^n}{2} \int_0^{\infty} du e^{-\left(\frac{\gamma}{2}+D+i\Delta+in\omega_{\text{A}}/2\right)u} J_n\left[2A\sin\left(\frac{\omega_{\text{A}} u}{2}\right)\right],$$

$$d_n = \frac{2D\gamma}{(\gamma + D + i\omega_{\text{A}} n)^2 - D^2}\left(c_n + c_{-n}^*\right).$$

In the next step, we split the intensities from Eqs. (S2) into elastic and inelastic parts, according to Eq. (S3). The following subsections are devoted to deriving each component of the intensities.

## A. Elastic intensities

We start by calculating the elastic part of the intensity from the first beam. We substitute Eq. (S3a) into Eq. (S2a) and after using the definition of the filter function from the main text we obtain, upon rearranging terms with $s$ and $s'$

$$I_1^{(\mathrm{el})}(t) = \frac{\Omega^2\Gamma^2}{\gamma^2} \sum_{n,m} c_n c_m^* e^{i\omega_\mathrm{A}(n-m)t} \int_0^\infty ds e^{-[\Gamma - i(\omega_1-\omega_\mathrm{L}-\omega_\mathrm{A}n)]s} \int_0^\infty ds' e^{-[\Gamma + i(\omega_1-\omega_\mathrm{L}-\omega_\mathrm{A}m)]s}.$$

If we now integrate over $s(s')$ and group the sum over the expressions with indices $n(m)$, we will see that $I_1^{(\mathrm{el})}(t)$ is

$$I_1^{(\mathrm{el})}(t) = I_0 \frac{\Gamma^2}{\pi} \left| \sum_n \frac{c_n e^{i\omega_\mathrm{A}nt}}{\Gamma - i(\omega_1 - \omega_\mathrm{L} - \omega_\mathrm{A}n)} \right|^2, \tag{S4}$$

where $I_0 = \pi\Omega^2/\gamma^2$ is the standard RF intensity of a weakly excited system [2].

The elastic intensity of the second beam, $I_2^{(\mathrm{el})}(t)$, is calculated in the same way, replacing $t$ and $\omega_1$ from Eq. (S4) with $t - \tau_\mathrm{d}$ and $\omega_2$, respectively. Hence,

$$I_2^{(\mathrm{el})}(t) = I_0 \frac{\Gamma^2}{\pi} \left| \sum_n \frac{c_n e^{i\omega_\mathrm{A}n(t-\tau_\mathrm{d})}}{\Gamma - i(\omega_2 - \omega_\mathrm{L} - \omega_\mathrm{A}n)} \right|^2. \tag{S5}$$

Elastic-scattering contribution of the interference term is calculated via substituting Eq. (S3a) into Eq. (S2c), which, after rearranging terms with $s$ and $s'$, gives

$$\begin{aligned} I_\mathrm{int}^{(\mathrm{el})}(t) =& \frac{2\Omega^2\Gamma^2}{\gamma^2} \,\mathrm{Re}\, e^{i\delta\omega t + i\omega_\mathrm{L}\tau_\mathrm{d}} \sum_{n,m} c_n c_m^* e^{i\omega_\mathrm{A}(n-m)t} e^{i\omega_\mathrm{A}m\tau_\mathrm{d}} \\ & \times \int_0^\infty ds e^{-[\Gamma - i(\omega_1-\omega_\mathrm{L}-\omega_\mathrm{A}n)]s} \int_0^\infty ds' e^{-[\Gamma + i(\omega_2-\omega_\mathrm{L}-\omega_\mathrm{A}m)]s'}, \end{aligned}$$

which after evaluating the integrals yields

$$I_\mathrm{int}^{(\mathrm{el})}(t) = 2I_0 \frac{\Gamma^2}{\pi} \,\mathrm{Re}\, e^{i\delta\omega t + i\omega_\mathrm{L}\tau_\mathrm{d}} \left[ \sum_n \frac{c_n e^{i\omega_\mathrm{A}nt}}{\Gamma - i(\omega_1 - \omega_\mathrm{L} - \omega_\mathrm{A}n)} \right] \left[ \sum_m \frac{c_m e^{i\omega_\mathrm{A}m(t-\tau_\mathrm{d})}}{\Gamma - i(\omega_2 - \omega_\mathrm{L} - \omega_\mathrm{A}m)} \right]^*. \tag{S6}$$

## B. Inelastic intensities

In order to calculate the inelastic intensities, we use $G_\mathrm{inel}(t_2, t_1) = G_\mathrm{inel}^*(t_1, t_2)$. Applying this property in (S2a) and using the definition of the filter function we obtain

$$\begin{aligned} I_1^{(\mathrm{inel})}(t) =& \Gamma^2 \int_0^\infty ds \int_0^s ds' e^{-\Gamma(s+s')} e^{i\omega_1(s-s')} G_\mathrm{inel}(t-s, t-s') \\ & + \Gamma^2 \int_0^\infty ds \int_s^\infty ds' e^{-\Gamma(s+s')} e^{i\omega_1(s-s')} G_\mathrm{inel}^*(t-s', t-s). \end{aligned} \tag{S7}$$

For the second part from Eq. (S7) we use the fact that for any function $f(s, s')$

$$\int_0^\infty ds \int_s^\infty ds' f(s,s') = \int_0^\infty ds \int_0^s ds' f(s', s)$$

and then $I_1^{\mathrm{inel}}(t)$ yields

$$I_1^{(\mathrm{inel})}(t) = 2\Gamma^2 \,\mathrm{Re} \int_0^\infty ds \int_0^s ds' e^{-\Gamma(s+s')} e^{i\omega_1(s-s')} G_\mathrm{inel}(t-s, t-s').$$

Then we substitute Eq. (S3b) and get

$$\begin{aligned} I_1^{(\mathrm{inel})}(t) =& \frac{\Omega^2\Gamma^2}{\gamma^2} \,\mathrm{Re} \sum_{n,m,k} b_m^* b_{n-k} d_k e^{i\omega_\mathrm{A}(n-m)t} \\ & \times \int_0^\infty ds e^{-\left[\Gamma + \frac{\gamma}{2} + D - i(\omega_1 - \omega_\mathrm{L} + \Delta - \omega_\mathrm{A}n)\right]s} \int_0^s ds' e^{-\left[\Gamma - \frac{\gamma}{2} - D + i(\omega_1 - \omega_\mathrm{L} + \Delta - \omega_\mathrm{A}m)\right]s'}. \end{aligned}$$

In the next step, we calculate the integral over $s'$, then over $s$ and finally, after rearranging terms, we obtain

$$I_1^{(\mathrm{inel})}(t) = I_0 \frac{\Gamma^2}{\pi} \operatorname{Re} \sum_{n,m,k} \frac{b_m^* b_{n-k} d_k}{\Gamma + \frac{\gamma}{2} + D - i(\omega_1 - \omega_{\mathrm{L}} + \Delta - \omega_{\mathrm{A}} n)} \frac{e^{i\omega_{\mathrm{A}}(n-m)t}}{2\Gamma + i\omega_{\mathrm{A}}(n-m)}. \tag{S8}$$

According to Eqs. (S2a) and (S2b), the inelastic intensity from the second input is found by replacing $t$ with $t - \tau_{\mathrm{d}}$ and $\omega_1$ with $\omega_2$ in Eq. (S8). Hence, we obtain

$$I_2^{(\mathrm{inel})}(t) = I_0 \frac{\Gamma^2}{\pi} \operatorname{Re} \sum_{n,m,k} \frac{b_m^* b_{n-k} d_k}{\Gamma + \frac{\gamma}{2} + D - i(\omega_2 - \omega_{\mathrm{L}} + \Delta - \omega_{\mathrm{A}} n)} \frac{e^{i\omega_{\mathrm{A}}(n-m)(t-\tau_{\mathrm{d}})}}{2\Gamma + i\omega_{\mathrm{A}}(n-m)}. \tag{S9}$$

The inelastic part of the interference term, after using properties of $G_{\mathrm{inel}}(t_1, t_2)$ and the filter definition, is

$$\begin{aligned} I_{\mathrm{int}}^{(\mathrm{inel})}(t) =& 2\Gamma^2 \operatorname{Re} e^{i\delta\omega t} \int_0^\infty ds' \int_0^{\tau_{\mathrm{d}}+s'} ds e^{-\Gamma(s+s')} e^{i(\omega_1 s - \omega_2 s')} G_{\mathrm{inel}}^*(t - \tau_{\mathrm{d}} - s', t - s) \\ &+ 2\Gamma^2 \operatorname{Re} e^{i\delta\omega t} \int_0^\infty ds' \int_{\tau_{\mathrm{d}}+s'}^\infty ds e^{-\Gamma(s+s')} e^{i(\omega_1 s - \omega_2 s')} G_{\mathrm{inel}}(t - s, t - \tau_{\mathrm{d}} - s'), \end{aligned}$$

which, after substituting Eq. (S3b) and rearranging terms with $s$ and $s'$, yields

$$\begin{aligned} I_{\mathrm{int}}^{(\mathrm{inel})}(t) =& \frac{\Omega^2\Gamma^2}{\gamma^2} \operatorname{Re} e^{i\delta\omega t} e^{-\left[\frac{\gamma}{2} + D - i(\omega_{\mathrm{L}} - \Delta)\right]\tau_{\mathrm{d}}} \sum_{n,m,k} b_m b_{n-k}^* d_k^* e^{i\omega_{\mathrm{A}} n \tau_{\mathrm{d}} - i\omega_{\mathrm{A}}(n-m)t} \\ &\times \int_0^\infty ds' e^{-\left[\Gamma + \frac{\gamma}{2} + D + i(\omega_2 - \omega_{\mathrm{L}} + \Delta - \omega_{\mathrm{A}} n)\right]s'} \int_0^{s'+\tau_{\mathrm{d}}} ds e^{-\left[\Gamma - \frac{\gamma}{2} - D - i(\omega_1 - \omega_{\mathrm{L}} + \Delta - \omega_{\mathrm{A}} m)\right]s} \\ &+ \frac{\Omega^2\Gamma^2}{\gamma^2} \operatorname{Re} e^{i\delta\omega t} e^{\left[\frac{\gamma}{2} + D + i(\omega_{\mathrm{L}} - \Delta)\right]\tau_{\mathrm{d}}} \sum_{n,m,k} b_m^* b_{n-k} d_k e^{i\omega_{\mathrm{A}} m \tau_{\mathrm{d}} + i\omega_{\mathrm{A}}(n-m)t} \\ &\times \int_0^\infty ds' e^{-\left[\Gamma - \frac{\gamma}{2} - D + i(\omega_2 - \omega_{\mathrm{L}} + \Delta - \omega_{\mathrm{A}} m)\right]s'} \int_{s'+\tau_{\mathrm{d}}}^\infty ds e^{-\left[\Gamma + \frac{\gamma}{2} + D - i(\omega_1 - \omega_{\mathrm{L}} + \Delta - \omega_{\mathrm{A}} n)\right]s}. \end{aligned}$$

In the next step, we calculate the integral over $s'$, then over $s$ and finally, after rearranging terms, we obtain

$$\begin{aligned} I_{\mathrm{int}}^{(\mathrm{inel})}(t) =& I_0 \frac{\Gamma^2}{\pi} \operatorname{Re} \sum_{n,m,k} e^{i\delta\omega t} \\ &\times \left( \frac{b_m b_{n-k}^* d_k^* e^{-\left[\frac{\gamma}{2} + D - i(\omega_{\mathrm{L}} - \Delta + \omega_{\mathrm{A}} n)\right]\tau_{\mathrm{d}}}}{\Gamma - \frac{\gamma}{2} - D - i(\omega_1 - \omega_{\mathrm{L}} + \Delta - \omega_{\mathrm{A}} m)} \frac{e^{-i\omega_{\mathrm{A}}(n-m)t}}{\Gamma + \frac{\gamma}{2} + D + i(\omega_2 - \omega_{\mathrm{L}} + \Delta - \omega_{\mathrm{A}} n)} \right. \\ &- \frac{b_m b_{n-k}^* d_k^* e^{-[\Gamma - i\omega_1 - i\omega_{\mathrm{A}}(n-m)]\tau_{\mathrm{d}}}}{\Gamma - \frac{\gamma}{2} - D - i(\omega_1 - \omega_{\mathrm{L}} + \Delta - \omega_{\mathrm{A}} m)} \frac{e^{-i\omega_{\mathrm{A}}(n-m)t}}{2\Gamma + i(\omega_2 - \omega_1) - i\omega_{\mathrm{A}}(n-m)} \\ &\left. + \frac{b_m^* b_{n-k} d_k e^{-[\Gamma - i\omega_1 + i\omega_{\mathrm{A}}(n-m)]\tau_{\mathrm{d}}}}{\Gamma + \frac{\gamma}{2} + D - i(\omega_1 - \omega_{\mathrm{L}} + \Delta - \omega_{\mathrm{A}} n)} \frac{e^{i\omega_{\mathrm{A}}(n-m)t}}{2\Gamma + i(\omega_2 - \omega_1) + i\omega_{\mathrm{A}}(n-m)} \right). \end{aligned} \tag{S10}$$

### C. Approximations

Since $\omega_{\mathrm{L}}\tau_{\mathrm{d}} \sim 1$, while $\omega_{\mathrm{A}}, |\omega_{\mathrm{L}} - \omega_1| << \omega_{\mathrm{L}}$, we can neglect $\omega_{\mathrm{A}}\tau_{\mathrm{d}}$ and assume $\omega_{\mathrm{L}}\tau_{\mathrm{d}} \approx \omega_1\tau_{\mathrm{d}}$ in Eqs. (S5), (S6), (S9), (S6) and Eq. (S10). Then the intensities are

$$
\begin{aligned}
I_{1,2}(t) =& I_0 \frac{\Gamma^2}{\pi} \left| \sum_n \frac{c_n e^{i\omega_{\mathrm{A}} n t}}{\Gamma - i(\omega_{1,2} - \omega_{\mathrm{L}} - \omega_{\mathrm{A}} n)} \right|^2 \\
&+ I_0 \frac{\Gamma^2}{\pi} \operatorname{Re} \sum_{n,m} \frac{b_m^* \sum_k b_{n-k} d_k}{\Gamma + \frac{\gamma}{2} + D - i(\omega_{1,2} - \omega_{\mathrm{L}} + \Delta - \omega_{\mathrm{A}} n)} \frac{e^{i\omega_{\mathrm{A}}(n-m)t}}{2\Gamma + i\omega_{\mathrm{A}}(n-m)},
\end{aligned} \tag{S11a}
$$

$$
\begin{aligned}
I_{\mathrm{int}}(t) =& I_0 \frac{\Gamma^2}{\pi} \operatorname{Re} \sum_{n,m} e^{i\delta\omega t + i\omega_{\mathrm{L}}\tau_{\mathrm{d}}} \\
&\times \left( \frac{2 c_n c_m^*}{\Gamma - i(\omega_1 - \omega_{\mathrm{L}} - \omega_{\mathrm{A}} n)} \frac{e^{i\omega_{\mathrm{A}}(n-m)t}}{\Gamma + i(\omega_2 - \omega_{\mathrm{L}} - \omega_{\mathrm{A}} m)} \right. \\
&+ \frac{b_m \sum_k b_{n-k}^* d_k^*}{\Gamma + \frac{\gamma}{2} + D + i(\omega_2 - \omega_{\mathrm{L}} + \Delta - \omega_{\mathrm{A}} n)} \frac{e^{-i\omega_{\mathrm{A}}(n-m)t}}{2\Gamma + i(\omega_2 - \omega_1) - i\omega_{\mathrm{A}}(n-m)} \\
&+ \left. \frac{b_m^* \sum_k b_{n-k} d_k}{\Gamma + \frac{\gamma}{2} + D - i(\omega_1 - \omega_{\mathrm{L}} + \Delta - \omega_{\mathrm{A}} n)} \frac{e^{i\omega_{\mathrm{A}}(n-m)t}}{2\Gamma + i(\omega_2 - \omega_1) + i\omega_{\mathrm{A}}(n-m)} \right).
\end{aligned} \tag{S11b}
$$

## 2. INTERFERENCE FRINGES PATTERN

In this part, we consider the geometry of the interference fringes. In order to derive their slant, we consider an approximation of a narrow bandwidth, $\Gamma \to 0$. It should be emphasized that the ideal filter is treated here as a mathematical abstraction to isolate the optical phase-dependent trajectories of the interference pattern. While a zero-bandwidth filter formally implies a loss of temporal resolution within the acoustic cycle (consistent with the time–frequency Heisenberg uncertainty principle), this formal regime allows us to extract the fundamental geometric backbone of the coherence.

We set the filters at selected lines of the RF spectrum, which means $\omega_{1,2} = \omega_{\mathrm{L}} + k_{1,2}\omega_{\mathrm{A}}$, where $(k_1, k_2)$ denotes the respective sideband orders. For the ideal filter, the inelastic parts of the intensities are suppressed. This is because the denominators of the inelastic terms retain a finite real part bounded by $(\Gamma + \gamma/2 + D) \to (\gamma/2 + D) > 0$, while within the numerator, we have $\Gamma \to 0$. Then, for the individual filtered intensities $I_1(t)$ and $I_2(t)$, only the elastic terms proportional to $|c_{k_1}|^2$ and $|c_{k_2}|^2$, respectively, survive, rendering them strictly time-independent constants. Consequently, the intensities for ideal filter are

$$
I_{1,2}(t) \approx I_{1,2}^{(\mathrm{el})}(t) \approx \frac{I_0}{\pi} |c_{k_{1/2}}|^2 \tag{S12a}
$$

$$
I_{\mathrm{int}}(t) \approx I_{\mathrm{int}}^{(\mathrm{el})}(t) \approx \frac{2I_0}{\pi} \operatorname{Re} c_{k_1} c_{k_2}^* e^{i\omega_{\mathrm{L}}\tau_{\mathrm{d}}} e^{i[\delta\omega + \omega_{\mathrm{A}}(k_1 - k_2)]t} \tag{S12b}
$$

The $I_1(t)$ and $I_2(t)$ intensities are time-independent. Hence, the maximal and minimal total intensities are here determined solely by Eq. (S12b), which leads to the relation

$$
\omega_{\mathrm{L}}\tau_{\mathrm{d}} = N\pi - \left[ k_1 - k_2 + \frac{\delta\omega}{\omega_{\mathrm{A}}} \right] \omega_{\mathrm{A}} t - \theta_{k_1,k_2}. \tag{S13}
$$

Even and odd N correspond to maximum and minimum intensity, respectively. We define here the phase shift $\theta_{k_1,k_2} = \arg\left(c_{k_1} c_{k_2}^*\right)$.

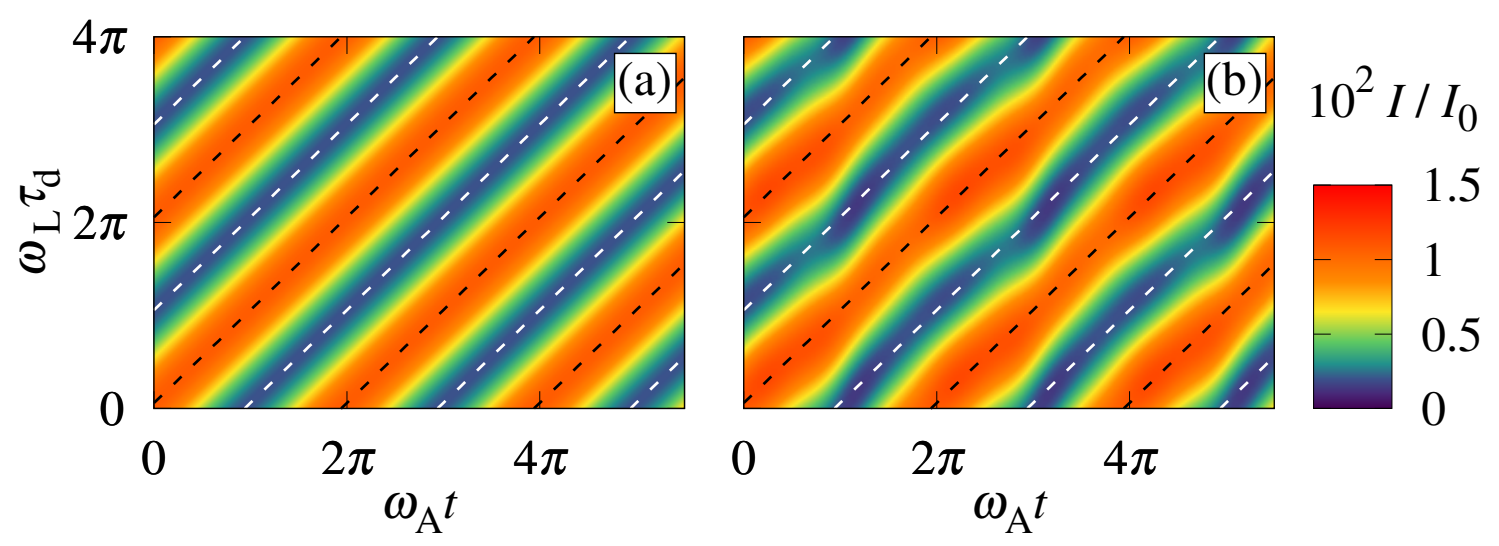


**Fig. S1.** Comparison of the coherent interference intensity between the zero phonon line (ZPL, $k_1 = 0$) and the first phonon sideband (PSB, $k_2 = 1$) under noise-free conditions ($D = 0$) and without an external frequency shift ($\delta\omega = 0$). (a) Ideal filter approximation ($\Gamma \to 0$) showing perfectly linear fringes with constant width. (b) Full result for a finite filter bandwidth ($\Gamma = 0.05\omega_A$) reproducing the periodic pulsation and envelope modulation discussed in the main text. In both panels, the analytical trajectories calculated directly from Eq. (S13) are superimposed as black and white dashed lines for the intensity maxima and minima, respectively. For $(k_1, k_2) = (0, 1)$ we have $\theta_{0,1} \approx -0.181$.

As shown in Fig. S1, the numerical results in the ideal filter approximation (Fig. S1(a)) exhibit perfectly linear fringes with a constant width. This comparison confirms that the periodic pulsation of the fringe thickness observed in Fig. S1b is a consequence of the finite filter bandwidth, which captures a broader spectral range. The geometry of the interferograms in both panels is in excellent agreement with the analytical prediction of Eq. (S13), validating the geometric ruler property across different filtering regimes.

## REFERENCES

1. R. A. Bogaczewicz and P. Machnikowski, "Precision of the acoustic control of single-photon scattering with semiconductor quantum dots," Opt. Lett. **50**, 888–891 (2025).
2. M. O. Scully and M. S. Zubairy, *Quantum Optics* (Cambridge University Press, 1997).


\clearpage

\end{document}